\documentclass[a4paper]{article}
\usepackage{authblk}
\usepackage[version=3]{mhchem}
\usepackage{graphicx} 
\usepackage{subfig}
\usepackage{multirow}
\usepackage{longtable}
\usepackage{achemso} 
\usepackage{array,booktabs}
\usepackage{float}
\usepackage{caption}
\usepackage{color}
\usepackage{enumerate}
\usepackage{amsmath}
\usepackage[none]{hyphenat}

\begin{document}

\title{Nitrogen Doped Graphene Quantum Dots as Possible Substrates to Stabilize Planar
Conformer of Au$_{20}$ Over its Tetrahedral Conformer: A Systematic DFT Study.}

\author[1]{Sharma SRKC Yamijala} 
\author[2]{Arkamita Bandyopadhyay}
\author[2,3,*]{Swapan K Pati \thanks{pati@jncasr.ac.in}} 
\affil[1]{Chemistry and Physics of Materials Unit, Jawaharlal Nehru Centre for Advanced Scientific Research, Bangalore 560064, India.}
\affil[2]{New Chemistry Unit, Jawaharlal Nehru Centre for Advanced Scientific Research, Bangalore 560064, India.}
\affil[3]{Theoretical Sciences Unit, Jawaharlal Nehru Centre for Advanced Scientific Research, Bangalore 560064, India.}
\affil[*]{Corresponding author}
\date{}

\maketitle


\begin{abstract}

Utilizing the strengths of nitrogen doped graphene quantum dot (N-GQD) as a substrate,
here in, we have shown that one can stabilize the catalytically more active planar
Au$_{20}$ (P-Au$_{20}$) compared to the thermodynamically more stable tetrahedral
structure (T-Au$_{20}$) on an N-GQD. Clearly, this simple route avoids the usage
of traditional transition metal oxide substrates which have been suggested and used
for stabilizing the planar structure for a long time. Considering the experimental
success in the synthesis of N-GQDs and in the stabilization of Au nanoparticles on
N-doped graphene, we expect our proposed method to stabilize planar structure will
be realized experimentally and will be useful for industrial level applications.

%
%

\end{abstract}

\noindent{Keywords:} Dimensionality crossover, Ab-initio calculations, Bi-layer graphene, Charge transfer, Catalysis. 

\section{Introduction}
\sloppy
Stability, ionization potential, electronic, magnetic, optical and catalytic
properties of gold clusters depend not only on their size but also on their
shape and charge state.\cite{cathrine_gold_book,Remacle_IJQC,Freund_PRB_2012_CaO,ayan_Au_clusters_IEEE_2011}
Stabilizing a particular conformer among the others, to achieve the desired
properties, is one of the active fields of research.
\cite{Landman_PRL_2006_MgO_on_Mo, Freund_PRL_2007_exp, Landman_JACS_2007_CO_oxidation,
Landman_PRL_2008_electric_field, Martinez_JPCC_2010_size_matters, Nisha_JACS,
Freund_angew_2011, Freund_JACS_2012_Cr_vs_Mo, Freund_PRB_2012_CaO, ayan_flytrap_2011}
When gold clusters are grown on a substrate, the nature of the substrate highly
dictates the stability of the conformer, and hence, also its shape. In the
past years, a large number of studies have been carried out on several substrates
mainly to understand the substrate properties in stabilizing a particular conformer
of the gold cluster. Many of these studies have concentrated on stabilizing the
catalytically active planar conformer of Au$_{20}$ cluster (P-Au$_{20}$) over the
thermodynamically stable tetrahedral conformer (T-Au$_{20}$)
\cite{ Landman_PRL_2006_MgO_on_Mo, Freund_PRL_2007_exp, Landman_JACS_2007_CO_oxidation,
Landman_PRL_2008_electric_field, Martinez_JPCC_2010_size_matters, Nisha_JACS,
Freund_angew_2011, Freund_JACS_2012_Cr_vs_Mo, Freund_PRB_2012_CaO}
and these studies have used metal oxides substrates, such as, MgO,
\cite{ Landman_PRL_2006_MgO_on_Mo, Freund_PRL_2007_exp,Freund_JACS_2012_Cr_vs_Mo,
Landman_JACS_2007_CO_oxidation, Landman_PRL_2008_electric_field, Nisha_JACS}
CaO
\cite{Freund_angew_2011, Freund_JACS_2012_Cr_vs_Mo, Freund_PRB_2012_CaO} etc.

	In gas phase, tetrahedral conformer is found to be the most stable conformer
both by experimental \cite{science_2003_Au20_tetra, science_2008_Gas_phase_au20}
and by several theoretical studies. \cite{Remacle_IJQC, Martinez_JPCC_2010_size_matters}
Same trend in the stability has been found even when Au$_{20}$ is on pristine MgO,
CaO substrates.\cite{Landman_PRL_2006_MgO_on_Mo, Freund_PRL_2007_exp,
Landman_JACS_2007_CO_oxidation, Landman_PRL_2008_electric_field,
Martinez_JPCC_2010_size_matters, Nisha_JACS, Freund_angew_2011,
Freund_JACS_2012_Cr_vs_Mo, Freund_PRB_2012_CaO}
Apart from its thermodynamic stability, T-Au$_{20}$ also has a larger HOMO-LUMO gap
(1.77 eV) compared to its other two-dimensional conformers.
\cite{Remacle_IJQC, Martinez_JPCC_2010_size_matters}
Thus, it is chemically more stable (or less reactive), and hence, not very useful for
catalytic applications. Less reactivity of T-Au$_{20}$ compared to P-Au$_{20}$ has
already been proved during the catalytic conversion of CO to CO$_{2}$ in the presence
of O$_{2}$ on a Mo-doped MgO substrate.\cite{Landman_JACS_2007_CO_oxidation}
Also, theoretical calculations have shown that both the electron accepting and
donating capabilities of P-Au$_{20}$ are more compared to that of T-Au$_{20}$
and such trend has been found to be common for planar clusters.
\cite{Martinez_JPCC_2010_size_matters}
Thus, to utilize the gold clusters in catalytic applications, it is required to
stabilize ``less stable but catalytically more active" conformers than the
``less reactive and thermodynamically more stable" conformers.
	
	Several previous works have shown different ways to stabilize the planar
conformer of Au$_{20}$.\cite{Landman_PRL_2006_MgO_on_Mo, Freund_PRL_2007_exp,
Landman_JACS_2007_CO_oxidation, Landman_PRL_2008_electric_field,
Martinez_JPCC_2010_size_matters, Nisha_JACS, Freund_angew_2011,
Freund_JACS_2012_Cr_vs_Mo,Freund_PRB_2012_CaO} Most of these works considered
metal-oxides as substrates and the methods used to to tune the morphology of
Au$_{20}$ include (i)  depositing thin metal-oxide films on transition metals
\cite{Landman_PRL_2006_MgO_on_Mo, Freund_PRL_2007_exp, Landman_JACS_2007_CO_oxidation,
Freund_PRB_2012_CaO}
(ii) application of external field \cite{Landman_PRL_2008_electric_field} when
depositing bulk metal-oxides on transition metals and
(iii) to add external dopants \cite{Nisha_JACS, Freund_angew_2011,
Freund_JACS_2012_Cr_vs_Mo, Freund_PRB_2012_CaO} to bulk metal-oxides without depositing
them on transition metals etc.
Unlike earlier works, in this study we have considered graphene quantum dots (GQDs),
\cite{sharma_GQD_JPCC_2013, sharma_GQD_PCCP_2013, sharma_X_shape_2014, Oded_Hod_GQDs_PRB_2008}
the zero-dimensional analogues of graphene, as substrate. We have considered
different possibilities like external doping by substituting the carbon atoms of GQD
with nitrogen or boron atoms, increasing the doping concentration, introduction of
defects, increasing the number of layers of GQDs etc. to see whether
we can stabilize P-Au$_{20}$ over T-Au$_{20}$ on GQDs. The main reason behind the
consideration of GQDs as substrate is mainly to due to a recent report by
Li et. al. on the successful preparation and stabilization of Pd nanoparticles (NPs)
on top of colloidal GQDs, \cite{Liang_JACS_2012_Pd_on_NGQDs} where their main focus
was on the Pd-carbon interaction. The same group (and also several other groups)
has also shown the successful synthesis of N-doped GQDs (NGQDs) with precise control
over the position of the dopant nitrogen. \cite{Liang_Accounts_2013_colloidal_GQDs,
RSC_adv_2012_gold_on_N_doped_graphene} Though, both experimental and theoretical works
exist on the interaction of Au clusters with N-doped graphene (not GQDs), they didn't
concentrate on tuning the morphology of Au clusters. In this work, we have shown that,
nitrogen doped GQDs can act as alternative substrates to doped metal-oxide substrates in
stabilizing the P-Au$_{20}$ over T-Au$_{20}$.

\begin{figure}
\center\includegraphics[scale=0.08]{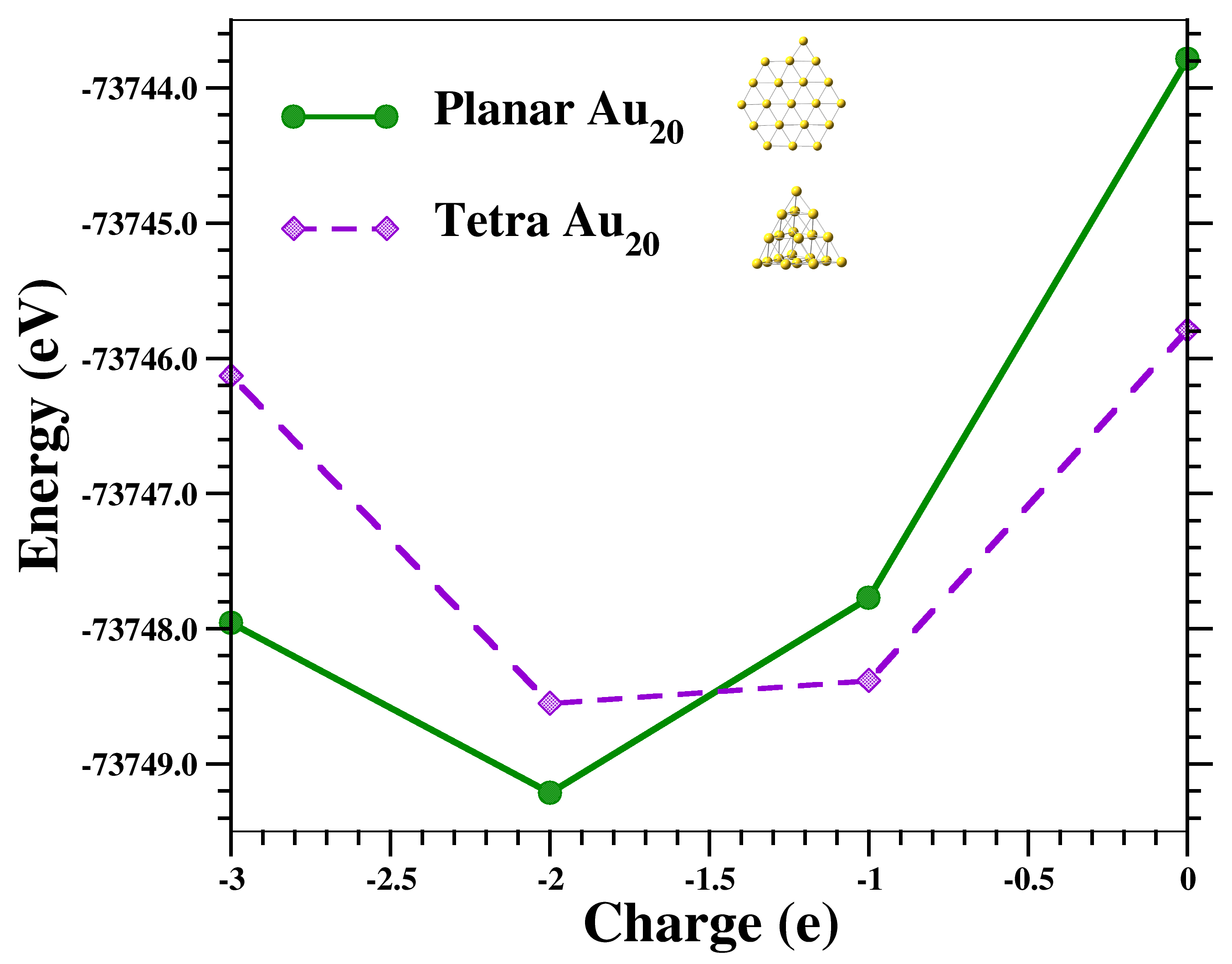}
\caption{Energy of P-Au$_{20}$ vs T-Au$_{20}$ as a function of charge. Observe the dimensionality cross-over at a charge of -2.}
\end{figure}
 
\section{Results and Discussions}
\sloppy

In previous works, \cite{Landman_PRL_2006_MgO_on_Mo, Freund_PRL_2007_exp,
Landman_PRL_2008_electric_field,Martinez_JPCC_2010_size_matters, Nisha_JACS,
Freund_angew_2011} it has been clearly mentioned that the main reason for the
stability of P-Au$_{20}$ over T-Au$_{20}$ on a doped metal-oxide substrate is due to
the greater charge transfer from the substrate to P-Au$_{20}$ than to T-Au$_{20}$
(and also due to the greater charge accumulation at the cluster-substrate interface).
But, these works didn't explain why there is a requirement of an oxide substrate if
charge transfer is the sole reason for the stability of P-Au$_{20}$. To address this
issue, first we have performed a series of calculations (at B3LYP/LANL2DZ level of
theory) on both P and T-Au$_{20}$ clusters with different charges (in gas phase)
and we find that the planar structure can be stabilized over tetrahedral structure
when charge on the system is -2 or more, as shown in figure 1 (see Supporting
Information (SI) to know the functional/basis-set dependency). Apart from this, in a
previous work, \cite{Nisha_JACS} it has been shown that P-Au$_{20}$ can be
stabilized over T-Au$_{20}$ on an Al-doped MgO substrate when substrate transfers
$\sim$ 0.9 e or more to the clusters. Thus, these results suggest that, even though a
substrate is not necessary to stabilize P-Au$_{20}$ over T-Au$_{20}$, it will help to
reduce the required amount of charge transfer in stabilizing P-Au$_{20}$.
To further prove the non-necessity of an oxide-substrate in stabilizing the planar
conformer, we have considered a single layer graphene quantum dot (GQD) as our
substrate and further calculations have been performed (at BLYP+DFT-D3/DZP level
of theory).

\begin{table}
\centering
\caption{Energy difference between P-Au$_{20}$ and T-Au$_{20}$ when they are isolated
and when they are on different substrates along with the energy of substrate cluster
interaction (E$_{SCI}$) is given for all the systems.}
\begin{tabular}{|c|c|c|c|} \hline

Systems        &   Energy  (eV)&   E$_{T}$ - E$_{P}$ (eV)  &  E$_{SCI}$ (eV) \\  \hline
P-Au$_{20}$         &   -17992.686  &           &           \\  \hline
T-Au$_{20}$         &   -17996.651  &   -3.965  &           \\  \hline
GQD            &   -35840.945  &           &	          \\  \hline
P-Au$_{20}$@GQD     &   -53840.212  & 	      &   -6.582  \\  \hline
T-Au$_{20}$@GQD     &   -53841.025  &   -0.813  &   -3.429  \\  \hline
N-GQD          &   -35955.676  &           &           \\  \hline
P-Au$_{20}$@N-GQD   &   -53955.573  &           &   -7.211  \\  \hline
T-Au$_{20}$@N-GQD   &   -53955.853  &   -0.280  &   -3.526  \\  \hline
2N-GQD         &   -36070.362  &           &           \\  \hline
P-Au$_{20}$@2N-GQD  &   -54070.726  &           &   -7.679  \\  \hline
T-Au$_{20}$@2N-GQD  &   -54070.701  &   0.025   &   -3.688  \\  \hline
3N-GQD         &   -36185.150  &           &           \\  \hline
P-Au$_{20}$@3N-GQD  &   -54185.571  &           &   -7.736  \\  \hline
T-Au$_{20}$@3N-GQD  &   -54185.454  &   0.117   &   -3.653  \\  \hline
4N-GQD         &   -36299.796  &           &           \\  \hline
P-Au$_{20}$@4N-GQD  &   -54300.703  &           &   -8.222  \\  \hline
T-Au$_{20}$@4N-GQD  &   -54300.194  &   0.509   &   -3.747  \\  \hline
5N-GQD         &   -36414.411  &           &           \\  \hline
P-Au$_{20}$@5N-GQD  &   -54415.489  &           &   -8.392  \\  \hline
T-Au$_{20}$@5N-GQD  &   -54414.844  &   0.645   &   -3.782  \\  \hline
6N-GQD         &   -36529.330  &           &           \\  \hline
P-Au$_{20}$@6N-GQD  &   -54530.402  &           &   -8.386  \\  \hline
T-Au$_{20}$@6N-GQD  &   -54529.753  &   0.649   &   -3.773  \\  \hline
B-GQD          &   -35762.547  &           &           \\  \hline
P-Au$_{20}$@B-GQD   &   -53761.879  &           &   -6.646  \\  \hline
T-Au$_{20}$@B-GQD   &   -53762.818  &   -0.939  &   -3.620  \\  \hline
pyN-GQD        &   -36030.984  &           &           \\  \hline
P-Au$_{20}$@pyN-GQD &   -54030.291  &           &   -6.621  \\  \hline
T-Au$_{20}$@pyN-GQD &   -54031.167  &   -0.876  &   -3.532  \\  \hline

\end{tabular}
\end{table}

In Table 1, we have given the energy difference (E$_{diff}$ = E$_{T}$ - E$_{P}$)
between the T-Au$_{20}$ and P-Au$_{20}$ clusters when they are isolated (i. e. not on
any substrate) and when they are on different substrates. Firstly, in accordance with
several previous studies, we find that the tetra conformer is more stable (negative
value of “E$_{T}$ - E$_{P}$”) than the planar conformer when the clusters are
isolated. We find the same trend even when the clusters are on a GQD substrate, although
the energy difference (E$_{diff}$) has reduced drastically (by $\sim$ 3 eV). In fact,
we find that this reduced E$_{diff}$ is due to the larger substrate-cluster
interaction (SCI) for the case of P-Au$_{20}$ than for T-Au$_{20}$, which in turn is
due to the shape of the P-Au$_{20}$ which allows all of its atoms to interact with
the substrate.  We have quantified the energy of SCI (E$_{SCI}$)as below:
E$_{SCI}$ = E$_{tot}$ - E$_{sub}$ - E$_{Au}$,
where, E$_{tot}$ is the total energy of the cluster on a substrate;
E$_{sub}$ and E$_{Au}$ are the energies of the isolated substrate and the Au cluster,
respectively, and the values are given in table 1. Clearly, E$_{SCI}$ for P-Au$_{20}$ is
$\sim$ 3 eV greater than the T-Au$_{20}$, when these clusters are on a GQD substrate.
Also, we find that (see table S1 of SI) there is $\sim$ 1 e charge transfer (CT) to
P-Au$_{20}$ from GQD, where as, it is only $\sim$ 0.2 e for T-Au$_{20}$. Thus, we
find that, though P-Au$_{20}$ has acquired higher amount of charge from GQD substrate
and has larger E$_{SCI}$ (when compared with T-Au$_{20}$), its stability is still less
than that of T-Au$_{20}$. This higher stability of T-Au$_{20}$ on a GQD substrate is
some what similar to what previously has been observed for the cases of MgO and CaO
substrates \cite{Landman_PRL_2006_MgO_on_Mo, Freund_PRL_2007_exp,
Landman_PRL_2008_electric_field,Martinez_JPCC_2010_size_matters,
Nisha_JACS,Freund_angew_2011} suggesting that our choice of substrate is correct and
further necessary steps have to be taken in order to acquire the required stability
of P-Au$_{20}$. Among the several previously implemented techniques, we find doping
the substrate with electron rich species \cite{Freund_JACS_2012_Cr_vs_Mo,
Nisha_JACS, Freund_angew_2011} as one of the simple and successful
technique for stabilizing P-Au$_{20}$ over T-Au$_{20}$ and we have doped our GQD
substrates with nitrogen (N) atoms.

Doping GQDs with N atoms can be of several ways, for example, pyridinic,
pyrrolic, substitutional [replacing C with N] etc.
Experimental studies on gold clusters stabilized on N-doped graphene have shown that
\cite{RSC_adv_2012_gold_on_N_doped_graphene} (i) substitutional and pyrrolic (pyridinic)
doping leads to n-type (p-type) graphene and (ii) dopant nitorgen sites in an n-type
graphene serves as electron donors and gold clusters acts as electron acceptors.
To verify these results, we have optimized the gold clusters on both substitutionally
doped N-GQD and pyridinic N-GQD (pyN-GQD). In agreement with these results, we find a
decrease (increase) in the negative ``E$_{T}$ - E$_{P}$" value, compared to that of 
pristine GQD, when doping is substitutional (pyridinic).
As our main aim is to stabilize P-Au$_{20}$, i.e. to attain a positive
``E$_{T}$ - E$_{P}$" value, we have performed all our further calculations only with
substitutional doping. We have varied the doping concentration from 0.44 \%
(i. e. one N atom in 228 C atoms) to 2.63 \% (6 N atoms in 228 C atoms) and while doping
more than one nitrogen, we have considered the experimental results of N-doped graphene
\cite{science_2011_visualilzing_N_in_graphene} and doped only the carbon atoms belonging
to same sub-lattice.

	In Table 1, we have given ``E$_{T}$ - E$_{P}$" values for all 
the different concentrations considered. Clearly, the most stable conformer of Au$_{20}$
on a GQD substrate has changed from tetra to planar for all the nitrogen dopant
concentrations greater than $\sim$ 0.88 \% (for the present level of theory) and we find
an increase in the stability of P-Au$_{20}$ with the increase in the dopant
concentrations. This is an interesting result, because it proves the non-necessity of an
oxide substrate for stabilizing catalytically active P-Au$_{20}$ conformer. Many N-doped
GQDs and graphene sheets have been synthesized.
\cite{science_2011_visualilzing_N_in_graphene,RSC_adv_2012_gold_on_N_doped_graphene,
Liang_Accounts_2013_colloidal_GQDs,Liang_JACS_2012_Pd_on_NGQDs}
So, we checked the robustness of our result against (i) the dopant atoms position
(ii) number of GQD layers (iii) exchange-correlation functional. Firstly,
for 2.63 \% concentration, we find that P-Au$_{20}$ is, at least, $\sim$ 0.26 eV
more stable than T-Au$_{20}$, even when all the dopant atoms are in a single zigzag
line of a GQD (which is not a favorable way of doping
\cite{science_2011_visualilzing_N_in_graphene}). Next, as the experimentally produced
GQDs generally contain more than one layer, we have also considered bi-layered GQDs.
With an increase in the number of layers, we find that the stability of P-Au$_{20}$ 
has further increased for the same number of dopant N-atoms. For example, when
substituting with two nitrogen atoms stability of P-Au$_{20}$ has increased from
$\sim$ 0.026 eV to $\sim$ 0.1 eV when moved from monolayer GQD to bi-layer GQD.
Similarly, when substituted with six nitrogen atoms, the stability has raised by
$\sim$ 0.3 eV for bi-layer GQD. Finally, we have changed the exchange correlation
functional and found that the trend is still
maintained, although the amount of gain in the stability is less (see SI). Thus, based
on our results and on the available experimental methods for growing N-GQDs as well as
gold clusters on N-doped graphene, we conjecture that experimentalists would
find a dimensionality cross-over from T-Au$_{20}$ to P-Au$_{20}$ on N-GQDs.

	Finally, to know the possible catalytically active sites of Au$_{20}$ clusters,
when supported on a N-GQD, we have plotted the iso-surfaces of charge transfer
between N-GQD and Au$_{20}$ clusters as shown in figure 2. To plot the
iso-surfaces of charge transfer, total electron density of the composite system
(i.e. N-GQD + Au$_{20}$) has been subtracted from the total electron density of the
substrate and the cluster with the same geometry (i.e. without any further optimization).
From these plots, it is clear that, major changes in the charge of the substrate occurred
only for the atoms which are below the Au$_{20}$ clusters. In the case of clusters,
major changes have occurred for the corner atoms than for the atoms which are in the
middle. Among T-Au$_{20}$ and P-Au$_{20}$ clusters, P-Au$_{20}$ has large number of
corner atoms and more number of atoms directly interacting with the substrate.
Also, we notice that the amount of charge accumulated at the substrate cluster interface
is more for P-Au$_{20}$ than for T-Au$_{20}$. Finally, for T-Au$_{20}$, only those atoms
which are directly above the N-GQD substrate have acquired more negative charge compared
to the ones in the upper layers. Thus, based on all these results and earlier reports
\cite{Landman_PRL_2006_MgO_on_Mo, Landman_PRL_2008_electric_field, Nisha_JACS}
 we expect that corner atoms of both the clusters will act as active sites for catalytic
applications and between P-Au$_{20}$ and T-Au$_{20}$, the former with more active sites
should be catalytically more active than T-Au$_{20}$.

\begin{figure}
\center\includegraphics[scale=0.5]{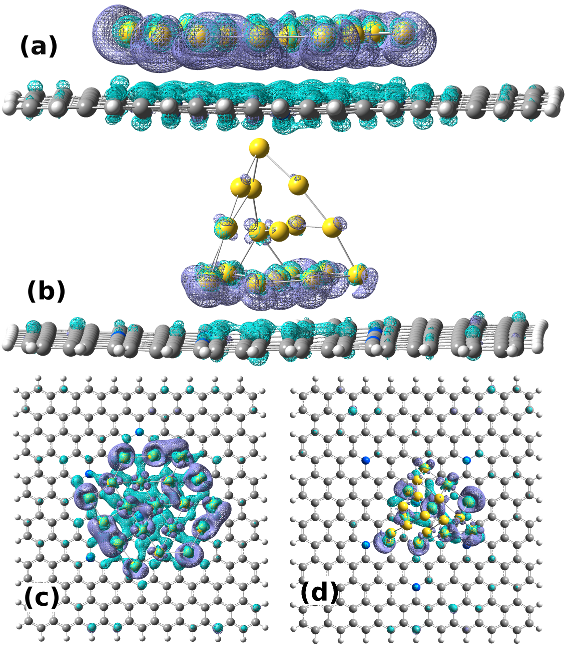}
\caption{Isosurface contours depecting the charge transfer process from substrate to 
Au$_{20}$ clusters. Top, bottom views of P-Au$_{20}$ are shown in (a), (c) and of
T-Au$_{20}$ in (b), (d). Iso-value of 0.001 e/\AA$^{3}$ is used for all the plots. Cyan
color depicts loss in electron density.}
\end{figure}

\section{Conclusions}
In conclusion, motivated by the recent successful synthesis of colloidal GQDs and N-GQDs
with precise control over the number of atoms, position of the dopants and their
application in stabilizing Pd nanoparticles, we have investigated several possibilities
of utilizing these doped/un-doped GQDs to stabilize the catalytically more useful
P-Au$_{20}$ compared to the thermodynamically more stable T-Au$_{20}$. Both
single-layer and bi-layer GQDs, with and without nitrogen dopants, have been considered
and we find that binding energy of P-Au$_{20}$ towards GQD is more ($\sim$ 3 eV)
compared to T-Au$_{20}$ and it is much more when the GQDs are doped with nitrogen and is
even more when the GQDs are bi-layered. Different concentrations of nitrogen doping 
have been considered and we find that, P-Au$_{20}$ can be stabilized over
T-Au$_{20}$, thermodynamically, by $\sim$ 1 eV when the N-dopant concentration is
$\sim$ 1.3 \% (i.e. 1N-atom for every 76-C atoms) in bi-layer GQDs. Also, from charge
transfer plots, we find that P-Au$_{20}$ has more active sites for catalysis. The main
point is that stronger interaction of P-Au$_{20}$ with N-GQD compared to
T-Au$_{20}$ is due to its large contact area with N-GQD substrate and also its ability
to accept more electrons.

\section{Computational Details}
\sloppy
Previous studies on the interaction between graphene and transition metal clusters
suggests that dispersion forces are important to exactly mimic the interaction between
gold and graphene and these studies have also shown that the empirical dispersion
correction i. e. DFT-D3 is sufficient to reproduce the results obtained with
the best methods (EE+vdW for Au-graphene; EE+vdW, M06-2X and MP2 calcuhide date in latexlations for
Au-coronene interactions) described in these works for graphene and gold interaction.
\cite{Granatier_JCTC_2011} We have performed all the calculations using
spin-unrestricted density functional theory with Becke-Lee-Yang-Parr (BLYP) GGA
exchange-correlation functional, \cite{becke_exchange_PRB_1988, LYP_correlation_PRB_1988}
along with Grimme's DFT-D3 dispersion correction,\cite{Grimme_DFT_D3_JCP_2010} as
implemented in the QUICKSTEP module of the CP2K package \cite{Hutter_quickstep_cpc_2005}
(unless otherwise mentioned explicitly). We have used the norm-conserving
Goedecker\textendash Teter\textendash Hutter (GTH) pseudopotentials, \cite{Hutter_GTH_PPs_PRB_1998,
Hutter_GTH_PPs_first_PRB_1996, Krack_GTH_PPs_TCA_2005} which are optimized in CP2K
package to use them along with the BLYP functional. CP2K uses a hybrid Gaussian and
plane wave method for the electronic representation.\cite{Hutter_GPW_Mol_physics_1997}
In this work, Kohn-Sham valence orbitals have been expanded using double
zeta valence polarized basis sets which are optimized for the GTH psuedopotentials
(DZVP\textendash MOLOPT\textendash SR\textendash GTH). Together with the NN50 
smoothing method, a 320 Ry density cut-off is used for the auxiliary basis set of
plane waves. To avoid any unwanted interaction with the periodic images, we have
considered a 38 $\times$ 38 $\times$ 38 $\AA$ cubic unit cell along with the poisson
\cite{Goedecker_poisson_JCP_2006, Goedecker_poisson_JCP_2007} solver (to ensure the
non existence of wave function after the edges of the simulation box). Geometry
optimizations have been performed using BFGS method and systems are optimized till
the force on each atom is less than 0.0001 Hartree/Bohr. G09 package
\cite{g09_reference} has been used 
to perform all the calculations on isolated gold clusters using different
exchange-correlation functionals, namely, PBE, BLYP, B3LYP and M06-2X with LANL2DZ 
basis set and LANL2 pseudopotentials.

\bibliography{au20_write_up}{}

\newpage

\begin{figure}
\centerline{\huge{\bf {Graphical TOC entry}}}
\center\includegraphics[scale=1.2]{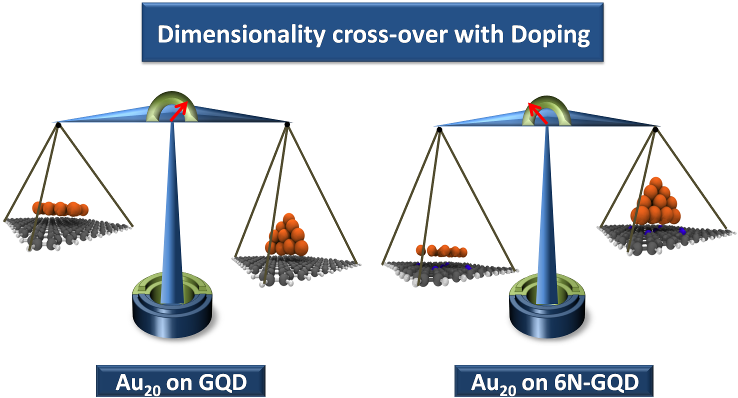}
\caption*{P-Au$_{20}$ becomes more stable than that of T-Au$_{20}$ after doping GQD substrate with nitrogen atoms.}
\end{figure}

\section{Supporting Information}
Additional computational details, table containing charges on individual species for
all the systems, charge dependency on exchange-correlation functional are given.

\newpage


\subsection{Charge required to observe a dimensionality crossover as a function of
exchange-correlation functional}

As shown in Figure S1, the amount of the charge required to obtain the dimensionality
crossover depends on the exchange-correlation functional (Exc) used. For BLYP and B3LYP
it is between -1 e and -2 e and for PBE and M06-2X it is between -2 and -3 e. Though the
amount of charge required varies with Exc, it is clear that, above a particular charge
dimensionality crossover will surely occur. Also, as mentioned in the main article, this
amount will change when the clusters are kept on a substrate.
  
\begin{figure}
\center\includegraphics[scale=0.15]{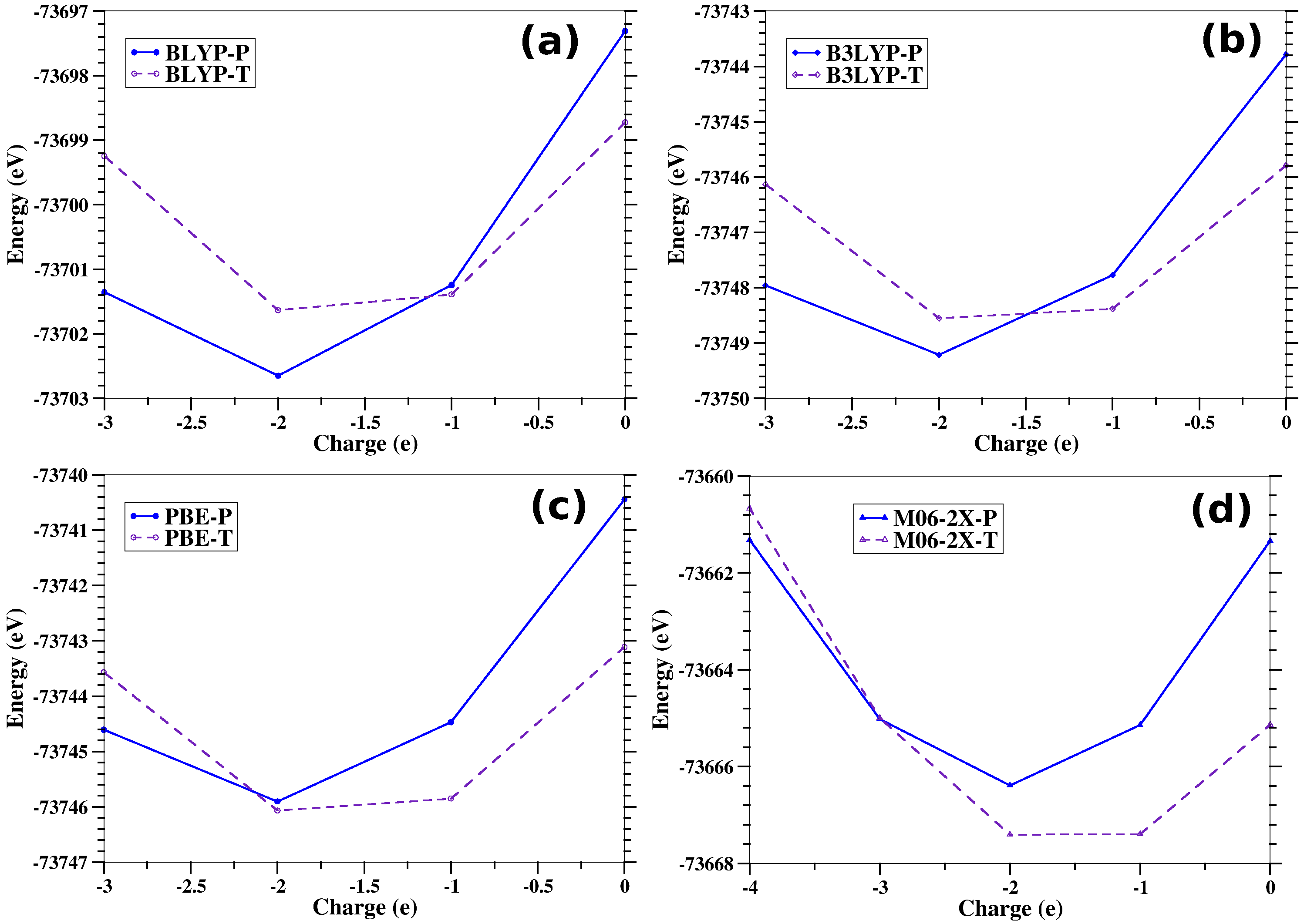}
\caption*{Figure S1: Energy of isolated Au$_{20}$ clusters as a function of charge and
exchange-correlation functionals. (a) BLYP (b) B3LYP (c) PBE and (d) M06-2X functionals.
P and T in the legends after the functional name denotes planar and tetra conformers,
respectively, of Au$_{20}$}.
\end{figure}

\subsection{Robustness check for stability of P-Au$_{20}$ over T-Au$_{20}$ on N-doped
GQDs with PBE functional}

From our calculations using PBE functional (by keeping all the other parameters like
cutoff, box length etc. of the BLYP functional), we find that P-Au$_{20}$ is less stable
than T-Au$_{20}$ on both monolayer and bi-layer GQDs even when six carbon atoms are
replaced with nitrogen atoms (though the difference has reduced to as less as $\sim$ 0.1 eV).
But, with tri-layer GQDs, we find that P-Au$_{20}$ has more stability than T-Au$_{20}$
(by $\sim$ 0.03 eV) when  tri-layer GQDs are doped with six nitrogen atoms. Further
calculations with higher concentrations are under progress and will be published else
where. Thus, our results are robust against change in the Exc and from these results we
conjecture that experimentalists would soon realize P-Au$_{20}$ clusters on N-doped few
layer GQDs. 

\begin{table}
\centering
\caption{Charges on the individual atoms (M$\ddot{u}$lliken charges) in the respective
systems. Amount of the charge transferred to the gold clusters from the substrates can
be directly identified by seeing column 3 (GOLD). Gain/loss of electron charge can be
seen by comparing the respective systems with the isolated systems. For example, by
comparing P-Au$_{20}$@GQD with GQD and P-Au$_{20}$, we can notice that, carbon has lost $\sim$ 0.95 e
charge and the same has been gained by gold.}
\begin{tabular}{|c|c|c|c|c|c|} \hline

Systems           &     CARBON   &    GOLD   &   NITROGEN   &    BORON   &   HYDROGEN  \\  \hline
P-Au$_{20}$            &      0.00    &    0.00   &     0.00     &     0.00   &     0.00    \\  \hline
T-Au$_{20}$            &      0.00    &    0.00   &     0.00     &     0.00   &     0.00    \\  \hline
GQD               &     -1.75    &    0.00   &     0.00     &     0.00   &     1.75    \\  \hline
P-Au$_{20}$@GQD        &     -0.83    &   -0.95   &     0.00     &     0.00   &     1.79    \\  \hline
T-Au$_{20}$@GQD        &     -1.58    &   -0.18   &     0.00     &     0.00   &     1.76    \\  \hline
N-GQD             &     -1.71    &    0.00   &    -0.03     &     0.00   &     1.74    \\  \hline
P-Au$_{20}$@N-GQD      &     -0.75    &   -1.07   &     0.04     &     0.00   &     1.78    \\  \hline
T-Au$_{20}$@N-GQD      &     -1.59    &   -0.20   &     0.04     &     0.00   &     1.75    \\  \hline
2N-GQD            &     -1.69    &    0.00   &    -0.05     &     0.00   &     1.73    \\  \hline
P-Au$_{20}$@2N-GQD     &     -0.42    &   -1.43   &     0.07     &     0.00   &     1.79    \\  \hline
T-Au$_{20}$@2N-GQD     &     -1.45    &   -0.34   &     0.04     &     0.00   &     1.75    \\  \hline
3N-GQD            &     -1.65    &    0.00   &    -0.08     &     0.00   &     1.73    \\  \hline
P-Au$_{20}$@3N-GQD     &     -0.27    &   -1.62   &     0.10     &     0.00   &     1.79    \\  \hline
T-Au$_{20}$@3N-GQD     &     -1.01    &   -0.79   &     0.05     &     0.00   &     1.76    \\  \hline
4N-GQD            &     -1.62    &    0.00   &    -0.11     &     0.00   &     1.72    \\  \hline
P-Au$_{20}$@4N-GQD     &     -0.29    &   -1.63   &     0.15     &     0.00   &     1.78    \\  \hline
T-Au$_{20}$@4N-GQD     &     -0.99    &   -0.83   &     0.07     &     0.00   &     1.75    \\  \hline
5N-GQD            &     -1.58    &    0.00   &    -0.13     &     0.00   &     1.71    \\  \hline
P-Au$_{20}$@5N-GQD     &     -0.25    &   -1.66   &     0.15     &     0.00   &     1.77    \\  \hline
T-Au$_{20}$@5N-GQD     &     -0.93    &   -0.87   &     0.05     &     0.00   &     1.75    \\  \hline
6N-GQD            &     -1.53    &    0.00   &    -0.18     &     0.00   &     1.71    \\  \hline
P-Au$_{20}$@6N-GQD     &     -0.20    &   -1.71   &     0.15     &     0.00   &     1.76    \\  \hline
T-Au$_{20}$@6N-GQD     &     -0.90    &   -0.86   &     0.02     &     0.00   &     1.74    \\  \hline
B-GQD             &     -1.63    &    0.00   &     0.00     &    -0.14   &     1.76    \\  \hline
P-Au$_{20}$@B-GQD      &     -0.77    &   -0.98   &     0.00     &    -0.05   &     1.80    \\  \hline
T-Au$_{20}$@B-GQD      &     -1.41    &   -0.35   &     0.00     &    -0.02   &     1.78    \\  \hline
pyN-GQD           &     -1.27    &    0.00   &    -0.49     &     0.00   &     1.76    \\  \hline
P-Au$_{20}$@pyN-GQD    &     -0.50    &   -1.02   &    -0.29     &     0.00   &     1.80    \\  \hline
T-Au$_{20}$@pyN-GQD    &     -1.07    &   -0.42   &    -0.28     &     0.00   &     1.77    \\  \hline

\end{tabular}
\end{table}

\subsection{Additional Computational Details}
\sloppy
It is important to mention that, BLYP calculations, in general, took more time to
converge compared to PBE calculations. Also, we find that energy fluctuations are huge
at the initial stages (some times even for 200 optimization steps) of geometry
optimization using BLYP. Though the  suggested remedies like decrease in the energy
gap or using FULL\textendash SINGLE\textendash INVERSE were helpful for monolayer
studies and bi-layer studies (although they also took huge time), they didn't help for 
tri-layer studies. Though an increase in plane wave cutoff (i. e. more than 320 Ry)
may solve the problem, we couldn't use it because of the issues in the memory (RAM)
while running the jobs (on our clusters).

\section{Acknowledgements}
S.S.R.K.C.Y., A.B. and S.K.P. acknowledge TUE-CMS, JNCASR for the computational
facilities. S.K.P. acknowledges DST for funding. A.B. acknowledges UGC for JRF.

\end{document}